\documentclass[12pt]{article}
\usepackage{amssymb,amsmath,epsfig}

\numberwithin{equation}{section}
\begin{document}

\title{\textbf{Fermions Tunneling and Quantum Corrections for
Quintessencial Kerr-Newman-AdS Black Hole}}

\author{Wajiha Javed \thanks{wajiha.javed@ue.edu.pk;
wajihajaved84@yahoo.com} and Rimsha Babar
\thanks{rimsha.babar10@gmail.com}\\
Department of Mathematics, University of Education,\\
Township, Lahore-54590, Pakistan.}

\date{}
\maketitle

\begin{abstract}
This paper is devoted to study charged fermion particles tunneling
through the horizon of Kerr-Newman-AdS black hole surrounded by quintessence
by using Hamilton-Jacobi ansatz. In our analysis, we investigate Hawking
temperature as well as quantum corrected Hawking temperature on account of
generalized uncertainty principle. Moreover, we discuss the effects of
correction parameter $\beta$ on the corrected Hawking temperature
$T_{e-H}$, graphically. We conclude that the temperature $T_{e-H}$
vanishes when $\beta=100$, whereas for $\beta<100$ and $\beta>100$, the
temperature turns out to be positive and negative, respectively. We
observe that the graphs of $T_{e-H}$ w.r.t. quintessence parameter
$\alpha$ exhibit behavior only for the particular ranges, i.e., $0<\alpha<1/6$, charge $0<Q\leq1$
and rotation parameter $0<a\leq1$. For smaller and larger values of negative
$\Lambda$, as horizon increases, the temperature decreases and
increases, respectively.
\end{abstract}

\section{Introduction}

Hawking radiation is the black-body radiation emitted by a black hole (BH) due to
quantum effects near the horizon of a BH \cite{C5}. It is named after the physicist
Stephen Hawking $(1979)$ who derived a theoretical argument for its
existence \cite{C6}. Hawking with his collaborators \cite{C8}
studied quantum mechanical uncertainty principle and observed that
the rotating BH should create and emit quantum particles. While, the base of BH
thermodynamics is propounded by Bekenstein \cite{C7}, who
predicted that the BH must have a finite entropy.

There are various methods to investigate the imaginary part of
outward radiated particles action. One of the
tunneling method is named as null geodesic method utilized by
Parik and Wilczek \cite{C9,C16}, which is
the extension of the analysis of Kraus and Wilczek \cite{C17}. They studied
the tunneling of massless scalar particles. For an outgoing
massive particle, the equation of motion is different from that of
massless particle. The trajectory of a massless particle is a null
geodesic. Another technique to investigate BH's tunneling process is
Hamilton-Jacobi ansatz introduced by Agheben et al. \cite{C18}, which
is the extension of complex route analysis of Padmanabhan et
al. \cite{C19}.

Later, Kerner and Mann \cite{C10} investigated tunneling process of
fermion particles by using Hamilton-Jacobi technique and obtained the
corresponding Hawking temperature. This phenomenon is based on
the calculations of imaginary part of action at horizon
which in turn is associated with the Boltzmann factor of emission at
Hawking temperature. Also, they \cite{C20} investigated fermions tunneling from
Rindler and typical non-rotating BH horizons. The WKB approximation is used to investigate the tunneling probability for
classically forbidden trajectory from interior to exterior region
through horizon. The expression of the tunneling probability $\Gamma$ is
given by
\begin{equation}
\Gamma=\exp[-\frac{2}{\hbar}ImI],
\end{equation}
where \emph{I} is the semi-classical action of the outgoing particle and $\hbar$ is Planck's constant.

Black holes are effective modes to explore the effects of quantum gravity
by investigating their thermodynamical properties. Using
generalized uncertainty principle (GUP) for BH physics, some thrilling implications
and consequences have been performed in literature \cite{C22}-\cite{C27}.
Nozari and Saghafi \cite{C25} discussed Hawking
radiation for massless scalar particles in the background geometry of
Schwarzschild BH by following the Parikh-Wilczek tunneling
technique and recovered the tunneling rate as well as corrected Hawking
temperature by considering GUP. Kerner et al. \cite{C31}, Li
et al. \cite{C32} and Jian et al. \cite{C33} investigated the
tunneling phenomenon of fermions from the Kerr and Kerr-Newman BHs
by applying the WKB approximation to the Dirac equation. Using Hamilton-Jacobi ansatz,
the GUP-deformed corrected Hawking temperatures for fermions are derived
for various curved spacetimes \cite{C37}-\cite{C45}.

By taking into account the quantum corrections and back-reaction effects, Singleton
et al. \cite{W1, W9} studied the information loss paradox, conservation of
energy and entropy. Banerjee and Majhi \cite{S1} studied the Hawking
radiation process by using quantum tunneling phenomenon at horizons
and investigated the back-reaction effects. The quantum tunneling
spectrum for scalar and fermion particles has also been discussed
in literature \cite{S2}-\cite{S6}.

Using Kerner and Mann's formulation, Sharif and Javed \cite{a4}
studied the fermions tunneling phenomenon to investigate Hawking temperatures for charged anti-de Sitter
BHs, charged torus-like BHs, $Pleba\acute{n}ski-Demia\acute{n}ski$
family of BHs, regular BHs and traversable wormholes. They \cite{C1} also
discussed Hawking radiation for a pair of
charged accelerating and rotating BHs involving NUT parameter.
Moreover, they \cite{A7} investigated quantum corrections for
regular BHs, i.e., Bardeen and ABGB BHs. Recently, Javed et al. \cite{[6]}
discussed charged vector particles tunneling for a pair of
accelerating and rotating BHs as well as for 5D gauged super-gravity
BHs.

During $1990$'s, it is observed that the universe is filled with matter and the
appealing power of gravity, which pulls all matter together. Later, the Hubble space
telescope perceptions for extremely far off supernovae demonstrated that quite a
while prior, the universe was really expending more gradually than it is today. So, the expansion
of universe has not been slow due to gravity, but it is accelerating. Eventually
scholars thought that possibly it was a consequence of Einstein's hypothesis of gravity, known as
cosmological constant $(\Lambda)$. Initially, $\Lambda$ is added by Einstein
as a steady term in his field equations of General Relativity in order to study
the static universe, later this effort was unsuccessful due to Hubble's
observations, which confirmed that our universe is expanding \cite{H1}.
Recently, $\Lambda$ is considered as a source term in the field equations,
which can be considered as mass of empty space or vacuum energy, adequately
utilizing dark energy to adjust gravity \cite{C28}.

The value of $\Lambda$ could be either positive or
negative as per background geometry. Dark energy is gravitationally repulsive, not attractive.
The nature of dark energy is still not well understood but
detected by its effects on the rate at which universe expand.
The idea of dark energy is more theoretical and numerous things are still remain
as matter of consideration \cite{C29}. Dark energy as a transient vacuum energy
resulting from the potential energy of a dynamical field, is
known as \textit{quintessence}. This form of dark energy varies in space
and time and distinguished from $\Lambda$. It
plays a vital role in the expansion theory of big bang \cite{C30}.

The description of dark energy in terms of negative cosmological constant (AdS space) has already been discussed in literature.
Xu and Wang \cite{B1} investigated
Kerr-Newman BH solution in the background of
quintessential field by utilizing the Newman-Janis algorithm. It is well
known fact that the Newman-Janis algorithm does not deal with the cosmological
constant, so they extended the solution of Kerr-Newman metric
to the Kerr-Newman-AdS solution surrounded by the quintessential dark energy through direct
computations satisfying Einstein Maxwell field equations in the background of quintessential
matter with negative cosmological constant. Also, they analyzed the singularity of Kerr-Newman AdS BH in the presence of
quintessential field which is similar as the case of Kerr BH.

In our analysis, the quintessential field has an equation of state
\begin{equation*}
p=\omega\rho,
\end{equation*}
where $\rho$ and $p$ denote the energy density and pressure, respectively.
While $\omega$ represents the state parameter having range,
$-1<\omega<-1/3$ \cite{B1}. For the given AdS BH which is surrounded by quintessential
field, their exist two cosmological horizons \cite{D1}. As $\omega\rightarrow-1$, the effects of
quintessence and cosmological constant will be similar \cite{D2}. Moreover, the $P–V$
criticality and thermodynamical properties (pressure, volume and Hawking
temperature) of quintessential RN-AdS BH is discussed by Li \cite{D3}.

In our analysis, we are going to extend the work of Xu and Wang in order to
investigate the thermodynamical properties of Kerr-Newman-AdS BH in the
presence of quintessential field for fermion particles. For this purpose, we have
considered quantum tunneling phenomenon of Hawking radiation to investigate
Hawking temperature at BH's horizon as well as its modified quantum corrected
form by utilizing Hamilton-Jacobi ansatz. Moreover, we investigate the effects
of quintessential field on BH's thermodynamics. The main purpose of
this paper is to investigate Hawking temperature from different aspects for AdS BH in the background of
quintessential field incorporating quantum effects. Moreover, we observe
the graphical behavior of corrected Hawking temperature with respect
to horizon ($r_{+}$) and quintessence parameter for correction parameter ($\beta$) and cosmological constant.

The paper is outlined as follows: In Section \textbf{2},
we introduce metric of the Kerr-Newman-AdS BH surrounded by the
quintessence. In Section \textbf{3}, we investigate the Hawking
radiation phenomenon for charged fermion particles for the above
mentioned BH and recoup the tunneling probability and Hawking
temperature. By utilizing the modified Dirac equation
incorporating GUP, the corrected Hawking temperature is determined in
section \textbf{4}, while section \textbf{5} provides
entropy corrections. Section \textbf{6} consists of the graphical
analysis of quantum corrected Hawking temperature, where
we study the effects of $\beta$ and $\Lambda$ in detail. The last section contains concluding remarks.

\section{Kerr-Newman-AdS BH with Quintessence}

The accelerating expansion of the universe implies the crucial
contribution of matter with negative pressure in the evolution of
universe. This expansion could also be the result of
cosmological constant or quintessence matter. If
quintessence matter exists all over the universe, it can also be
around a BH. Kerr BH has many interesting
properties distinct from its non-spinning counterpart, i.e., Schwarzschild BH. Newman
and Janis \cite{R1}-\cite{R4} analyzed that the Kerr metric \cite{R6} could
be obtained from the Schwarzschild metric using a complex
transformation within the framework of the Newman-Penrose formalism
\cite{R7}. A similar procedure was applied to the Reissner-Nordstrom
metric to generate Kerr-Newman metric \cite{R8}. The Newman-Janis algorithm proved to be prosperous in
generating new stationary solutions of the Einstein field equations \cite{R9}-\cite{R12}. Zhaoyi and Wang \cite{B1} derived
the Kerr-Newman-AdS solution in the presence of quintessence by using
Newman-Janis algorithm and complex computations.

The line-element can be written as \cite{B1}
\begin{eqnarray}
ds^{2}&=&\frac{\Sigma^2}{\Delta_{r}}dr^2+
\frac{\Sigma^2}{\Delta_{\theta}}d\theta^2
+\frac{\Delta_{\theta}\sin^2\theta}{\Sigma^2}
\left(a\frac{dt}{\Xi}-(r^2+a^2)\frac{d\phi}{\Xi}\right)^2\nonumber\\
&-&\frac{\Delta_{r}}{\Sigma^2}\left(\frac{dt}{\Xi}
-a\sin^2\theta\frac{d\phi}{\Xi}\right)^2,\label{B1}
\end{eqnarray}
where
\begin{eqnarray}
\Delta_{r}&=&R^2\left(1-\frac{r^2}{\ell^2}\right)+Q^2-2Mr-\alpha r^{1-3\omega}
,~~~\Delta_{\theta}=1+\frac{a^2\cos^2\theta}{\ell^2},\nonumber\\
R^2&=&r^2+a^2,~~~\ell^2=\frac{3}{\Lambda}
,~~~\Xi=1+\frac{a^2}{\ell^2},~~~\Sigma^2=r^2+a^2\cos^2\theta.\nonumber
\end{eqnarray}
In above expressions, $M$ is BH mass, $Q$ is BH charge defined as $Q^{2}=q_{e}^{2}+q_{m}^{2}$, while $q_{e}^{2}$ and
$q_{m}^{2}$ are being electric and magnetic charge parameters,
respectively, $a$ is the rotation parameter, $\ell$ is the curvature
radius represented by cosmological constant $\Lambda<0$, whereas
$\alpha$ is the quintessence parameter and the state parameter
$\omega$ ranges from $-1<\omega<-\frac{1}{3}$.

For $\Lambda=0$, the relationship between $\alpha$ and $\omega$ is
defined as follows \cite{B1}
\begin{equation}
\alpha\leq\frac{2}{(1-3\omega)}8^{\omega}.\label{B9}
\end{equation}
It is important to note that only for the fixed value of the state
parameter $\omega=-2/3$, one can obtain four
horizons, these are, inner, outer and two cosmological horizons $r_q$ (determined by the quintessence)
and $r_c$ (influenced by the cosmological constant).
For $\omega=-\frac{2}{3}$, the above expression (\ref{B9})
implies $\alpha\leq\frac{1}{6}$. For $\Lambda\neq0$, four roots can be
obtained by taking $\Delta_{r}=0$. For $\omega=-\frac{2}{3}$, the
horizon equation will become
\begin{equation}
r^{4}+\frac{3\alpha}{\Lambda}r^{3}+\left(a^{2}-\frac{3}{\Lambda}\right)r^{2}
+\frac{6M}{\Lambda}r-\frac{3}{\Lambda}\left(a^{2}+Q^{2}\right)=0.
\end{equation}
The above fourth order algebraic equation can be expressed in terms of
its roots, i.e.,
\begin{equation}
(r-r_{in})(r-r_{out})(r-r_{q})(r-r_{c})=0,
\end{equation}
where, $r_{-}$ is a cauchy (inner) horizon, $r_{+}$ is the event
(outer) horizon, $r_{q}$ and $r_{c}$ are two cosmological horizons.

The line-element (\ref{B1}) can be expressed as
\begin{equation}
ds^2=-f(r,\theta)dt^2+\frac{1}{g(r,\theta)}dr^2+\frac{1}{\rho(r,\theta)}
d\theta^2+K(r,\theta)d\phi^2
-2H(r,\theta)dtd\phi,
\end{equation}
where the metric functions are defined by
\begin{eqnarray*}
f(r,\theta)&=&\frac{-a^2\sin^2\theta\Delta_{\theta}+\Delta_{r}}{\Sigma^2\Xi^2},
~~~H(r,\theta)=\frac{a\sin^2\theta(-\Delta_{r}+R^2\Delta_{\theta})}{\Sigma^2\Xi^2},\\
K(r,\theta)&=&\frac{\sin^2\theta(\Delta_{\theta}R^4-a^2\sin^2\theta\Delta_{r})}{\Sigma^2\Xi^2},
~~~g(r,\theta)=\frac{\Delta_{r}}{\Sigma^2}
,~~~\rho(r,\theta)=\frac{\Delta_{\theta}}{\Sigma^2}.
\end{eqnarray*}
The electromagnetic vector potential $A_{\mu}$, is given by \cite{B38}
\begin{equation}
A_{\mu}=\frac{1}{\Sigma^2\Xi}\left[-(q_{e}r+aq_{m}\cos\theta)dt+
(a\sin^2\theta q_{e}r+q_{m}R^2\cos\theta)d\phi\right].\label{R1}
\end{equation}
The angular velocity is defined as \cite{a3}
\begin{equation}
\Omega=-\frac{g_{t\phi}}{g_{\phi\phi}}=\frac{a(R^2\Delta_{\theta}
-\Delta_{r})}{R^4\Delta_{\theta}-a^2\sin^2\theta\Delta_{r}},
\end{equation}
at horizon it can be expressed as \cite{C1}
\begin{equation}
\Omega_{H}=\frac{H(r_{+},\theta)}{K(r_{+},\theta)}=\frac{a}{r^2_{+}+a^2}.
\end{equation}
The inverse of $f(r,\theta)$ can be written as \cite{C1}
\begin{equation}
F(r,\theta)=f(r,\theta)+\frac{H^2(r,\theta)}{K(r,\theta)}=
\frac{\Delta_{r}\Delta_{\theta}\Sigma^{2}(r,\theta)}{\Xi^{2}
(R^4\Delta_{\theta}-a^2\sin^2\theta\Delta_{r})}.
\end{equation}

\section{Charged Fermions Tunneling}

This section is devoted to investigate massive charged fermions tunneling
phenomenon for Kerr-Newman-AdS BH surround by quintessence having
electric and magnetic charges. For this purpose, we will consider the covariant Dirac
equation, given as \cite{C1}
\begin{equation}
\iota\gamma^{\mu}\left(D_{\mu}-\frac{\iota q}{\hbar}A_{\mu}\right)\Psi
+\frac{m}{\hbar}\Psi=0,~~~\mu=0,1,2,3\label{C1}
\end{equation}
where $q$ is electric charge and $\Psi$ is wave function, while
\begin{eqnarray}
D_{\mu}=\partial_{\mu}+\Omega_{\mu},~~~\Omega_{\mu}=
\frac{1}{2}\iota\Gamma_{\mu}^{\alpha\beta}\Sigma_{\alpha\beta},
~~~\Sigma_{\alpha\beta}=\frac{1}{4}\iota[\gamma^{\alpha},\gamma^{\beta}].
\end{eqnarray}
Here, $\gamma^{\mu}$ satisfies the following identities $[\gamma^{\alpha},\gamma^{\beta}]=-[\gamma^{\beta},\gamma^{\alpha}]$
for $\alpha\neq\beta$ and $[\gamma^{\alpha},\gamma^{\beta}]=0$ for $\alpha=\beta$. Using these relationships as well as the symmetric property of connection symbol $\Gamma_{\mu}^{\alpha\beta}$, we can obtain $\Omega_{\mu}=0$, which yields $D_{\mu}=\partial_{\mu}$. Thus, the given Eq.(\ref{C1}) reduces to the following form \cite{C1,H2}
\begin{equation}
\iota\gamma^{\mu}\left(\partial_{\mu}-\frac{\iota q}{\hbar}A_{\mu}\right)\Psi
+\frac{m}{\hbar}\Psi=0,\label{C2}
\end{equation}
where Dirac matrices $\gamma^{\mu}$ are
\begin{eqnarray}
&&\gamma^{t}=\frac{1}{\sqrt{F(r,\theta)}}\gamma^{0},~~~\gamma^{r}
=\sqrt{g(r,\theta)}\gamma^{3},~~~\gamma^{\theta}=\sqrt{\rho(r,\theta)}\gamma^{1},\nonumber\\
&&\gamma^{\phi}=\frac{1}{\sqrt{K(r,\theta)}}\left(\gamma^{2}+\frac{H(r,\theta)}
{\sqrt{F(r,\theta)K(r,\theta)}}\gamma^{0}\right),\nonumber
\end{eqnarray}
here $\gamma^{c}$'s are for chiral and $\gamma$'s are for Minkowski space, defined as \cite{C31}
\begin{equation}
\gamma^{0}=\left({\begin{array}{cc}0 & I\\-I & 0\\ \end{array}}\right),\gamma^{1}
=\left({\begin{array}{cc}0 & \sigma^{1}\\\sigma^{1} & 0\\ \end{array}}\right),\gamma^{2}
=\left({\begin{array}{cc}0 & \sigma^{2}\\\sigma^{2} & 0\\ \end{array}}\right),\gamma^{3}
=\left({\begin{array}{cc}0 & \sigma^{3}\\\sigma^{3} & 0\\ \end{array}}\right).
\end{equation}
Here, Pauli sigma matrices $\sigma$'s are defined as
\begin{equation}
\sigma^{1}=\binom{0~~~~~1}{1~~~~~0},~~~\sigma^{2}=\binom{0~~-i}{i~~~~~0},
~~~\sigma^{3}=\binom{1~~~~0}{0~-1} \label{A3}.
\end{equation}
Spin particles have two types of spin states, spin-up and spin-down,
the corresponding wave functions are defined as, respectively
\begin{eqnarray}
\Psi_{\uparrow}(t,r,\theta,\phi)&=&\left[{\begin{array}{cccc}
A(t,r,\theta,\phi) \\ 0\\B(t,r,\theta,\phi) \\ 0\\ \end{array}}\right]
\exp\left[\frac{i}{\hbar}I_{\uparrow}(t,r,\theta,\phi)\right],\\
\Psi_{\downarrow}(t,r,\theta,\phi)&=&\begin{bmatrix}0\\
C(t,r,\theta,\phi)\\0\\D(t,r,\theta,\phi)
\end{bmatrix}\exp\left[\frac{i}{\hbar}I_{\downarrow}(t,r,\theta,\phi)\right],
\end{eqnarray}
where $I_{\uparrow}$ (spin-up) and $I_{\downarrow}$ (spin-down) represent particles action. Here, we discuss only the
spin-up case, the spin-down case is similar.
Moreover, the particles motion is considered in positive
radial direction and the corresponding action can be considered as \cite{C1}
\begin{equation}
I_{\uparrow}=-Et+J\phi+R(r,\theta),\label{C3}
\end{equation}
where $E$ and $J$ represent particles energy and angular
momentum, respectively, while $R$ is arbitrary function of $r$ and $\theta$. Using WKB
approximation and the ansatz (\ref{C3}) for spin-up particles into the
Dirac equation with $\iota A=B$ and $\iota B=A$, by applying the
Taylor's expansion near the event horizon, we can obtain the following set of
equations, i.e.,
\begin{eqnarray}
&-&B\left[\frac{-E+\Omega_{H}J-q\left(\frac{q_{e}r+aq_{m}}{(r^{2}_{+}
+a^{2})^{2}}\right)}{\sqrt{(r-r_{+})F_{r}(r_{+},\theta)}}+
\sqrt{(r-r_{+})g_{r}(r_{+},\theta)}R_{r}(r_{+},\theta)\right]\nonumber\\
&+&mA=0,\label{C4}\\&-&B\left[\frac{\iota}{\sqrt{K(r_{+},\theta)}}
\left(J+q\frac{a\sin^{2}\theta q_{e}r_{+}+
R^{2}_{+}q_{m}\cos\theta}{\Sigma^{2}(r_{+},\theta)}\right)\right.\nonumber\\
&+&\left.\sqrt{\rho(r_{+},\theta)}R_{\theta}(r,\theta)\right]=0,\label{C5}\\
&+&A\left[\frac{-E+\Omega_{H}J-q\left(\frac{q_{e}r+aq_{m}}{(r^{2}_{+}+a^{2})^{2}}\right)}
{\sqrt{(r-r_{+})F_{r}(r_{+},\theta)}}+
\sqrt{(r-r_{+})g_{r}(r_{+},\theta)}R_{r}(r_{+},\theta)\right]\nonumber\\
&+&mB=0,\label{C6}\\
&-&A\left[\frac{\iota}{\sqrt{K(r_{+},\theta)}}
\left(J+q\frac{a\sin^{2}\theta q_{e}r_{+}+R^{2}_{+}
q_{m}\cos\theta}{\Sigma^{2}(r_{+},\theta)}\right)\right.\nonumber\\
&+&\left.\sqrt{\rho(r_{+},\theta)}R_{\theta}(r,\theta)\right]=0.\label{C7}
\end{eqnarray}
In the above set of Eqs.(\ref{C4})-(\ref{C7}), the expressions can be
defined as
\begin{equation}
F_{r}(r_{+},\theta)=\frac{2\left[r_{+}(1-\frac{a^2}{\ell^2})-\frac{2r^3_{+}}
{\ell^2}-M-\frac{\alpha(1-3\omega)r^{-3\omega}_{+}}{2}\right]
(r^2_{+}+a^2\cos^2\theta)}{(r^2_{+}+a^2)^2},\nonumber
\end{equation}
\begin{eqnarray}
g_{r}(r_{+},\theta)&=&\frac{\Delta_{r}(r_{+})}{\Sigma^2(r_{+},\theta)}
=\frac{2\left[r_{+}(1-\frac{a^2}{\ell^2})-\frac{2r^3_{+}}{\ell^2}-M-
\frac{\alpha(1-3\omega)r^{-3\omega}_{+}}{2}\right]}{(r^2_{+}+a^2\cos^2\theta)}.\nonumber
\end{eqnarray}

For $m=0$, it is feasible to extract $\frac{1}{\sqrt{\Sigma(r_{+},\theta)}}$
from Eqs.(\ref{C4}) and (\ref{C6}), to make these equations independent
of $\theta$. Moreover, Eqs.(\ref{C5}) and (\ref{C7}) have no definite $r$
dependence. It is possible to separate the function $R$. So, near the horizon,
the arbitrary function $R(r,\theta)$ can be separated as follows
\begin{equation}
R(r,\theta)=R(r)+\Theta(\theta).
\end{equation}
Equations (\ref{C5}) and (\ref{C7}) provide the same equation for
$\Theta$ disregarding of $A$ or $B$. For $m=0$, Eqs.(\ref{C4}) and
(\ref{C6}) have two feasible solutions, given below
\begin{eqnarray}
R'(r_+)=R'_{+}(r_+)=\left[\frac{\left\{E-\Omega_{H}J-q\left(\frac{q_{e}r
+aq_{m}}{(r^{2}_{+}+a^{2})^{2}}\right)\right\}(r^{2}_{+}+a^{2})}
{\Delta_{r}(r_{+})(r-r_{+})}\right],\label{A1}\\R'(r_+)=R'_{-}(r_+)
=-\left[\frac{\left\{E+\Omega_{H}J-q\left(\frac{q_{e}r+aq_{m}}{(r^{2}_{+}+a^{2})^{2}}\right)\right\}
(r^{2}_{+}+a^{2})}{\Delta_{r}(r_{+})(r-r_{+})}\right],\label{A2}
\end{eqnarray}
where
\begin{equation*}
\Delta_{r}(r_{+})=2\left[r_{+}(1-\frac{a^2}{\ell^2})-\frac{2r^3_{+}}{\ell^2}
-M-\frac{\alpha(1-3\omega)r^{-3\omega}_{+}}{2}\right],
\end{equation*}
$R_{+}$ and $R_{-}$ represent the radial solution of outgoing and
incoming particles action, respectively.
Solving for $R_{+}(r_+)$, Eq.(\ref{A1}) implies
\begin{equation}
R_{+}(r_+)=\int\left[\frac{\left\{E-\Omega_{H}J-q
\left(\frac{q_{e}r+aq_{m}}{(r^{2}_{+}+a^{2})^{2}}\right)\right\}
(r^{2}_{+}+a^{2})}{\Delta_{r}(r_{+})(r-r_{+})}\right]dr.\nonumber
\end{equation}
After integrating the above expression at the pole, we get
\begin{equation}
R_{+}(r_+)=\pm\iota\pi\frac{\left[E-\Omega_{H}J-q\left(\frac{q_{e}r+
aq_{m}}{(r^{2}_{+}+a^{2})^{2}}\right)\right]
(r^2_{+}+a^2)}{2\left[r_{+}(1-\frac{a^2}{\ell^2})-\frac{2r^3_{+}}{\ell^2}-
M-\frac{\alpha(1-3\omega)r^{-3\omega}_{+}}{2}\right]}.\nonumber
\end{equation}
The above equation implies that
\begin{equation}
ImR_{+}=\pm\pi\frac{\left[E-\Omega_{H}J-q\left(\frac{q_{e}r+aq_{m}}
{(r^{2}_{+}+a^{2})^{2}}\right)\right]
(r^2_{+}+a^2)}{2\left[r_{+}(1-\frac{a^2}{\ell^2})-\frac{2r^3_{+}}{\ell^2}-
M-\frac{\alpha(1-3\omega)r^{-3\omega}_{+}}{2}\right]}.
\end{equation}

The tunneling probability in terms of spatial ($\oint p_{r}dr$) and temporal contribution ($\breve{E}\Delta t^{out,in}$) is
given as follows \cite{B4}-\cite{W4}
\begin{equation}
\Gamma=\exp\left[\frac{1}{\hbar}\left(Im(\breve{E}\Delta
t^{out})+Im(\breve{E}\Delta t^{in})-Im\oint
p_{r}dr\right)\right],\label{D1}
\end{equation}
where $\oint p_rdr=\int p^{+}_{r}dr-\int
p^{-}_{r}dr$, whereas $p^{\pm}_{r}=\pm\partial_{r}R$. The spatial contribution can be calculated as
\begin{eqnarray}
\Gamma_{spatial}&\propto&\exp\left[-\frac{1}{\hbar} Im\oint
p_{r}dr\right]\nonumber\\&=&\exp\left[-\frac{1}{\hbar}
Im\left(\int p^{+}_{r}dr-\int p^{-}_{r}dr \right)\right]\nonumber\\
&=&\exp\left[-\frac{\pi}{\hbar}\left(\frac{\breve{E}(r^2_{+}+a^2)}{r_{+}(1-\frac{a^2}{\ell^2})
-\frac{2r^3_{+}}{\ell^2}-M
-\frac{\alpha(1-3\omega)r^{-3\omega}_{+}}{2}}
\right)\right].
\end{eqnarray}
The connection between interior and exterior regions of a BH
defines the temporal part. For $t\rightarrow
t-\frac{\iota\pi}{(2\kappa)}$, we can define the temporal contribution as
\begin{equation}
Im(\breve{E}\Delta t^{out,in})=\frac{-\breve{E}\pi}{(2\kappa)},\nonumber
\end{equation}
where
\begin{equation}
\breve{E}=\left[E-\Omega_{H}j-q\left(\frac{q_{e}r+aq_{m}}
{(r^{2}_{+}+a^{2})^{2}}\right)\right]\nonumber
\end{equation}
and
\begin{equation*}
\kappa=\left[\frac{r_{+}(1-\frac{a^2}{\ell^2})
-\frac{2r^3_{+}}{\ell^2}-M
-\frac{\alpha(1-3\omega)r^{-3\omega}_{+}}{2}}{(r^2_{+}+a^2)}\right].
\end{equation*}
Thus, the total temporal rate can be obtained as
\begin{eqnarray}
\Gamma_{temp.}&\propto&\exp\left[\frac{1}{\hbar}\left(Im(\breve{E}\Delta
t^{out})+
Im(\breve{E}\Delta t^{in})\right)\right],\nonumber\\
&=&\exp\left[-\frac{\pi}{\hbar}\left(\frac{\breve{E}(r^2_{+}+a^2)}{r_{+}(1-\frac{a^2}{\ell^2})
-\frac{2r^3_{+}}{\ell^2}-M
-\frac{\alpha(1-3\omega)r^{-3\omega}_{+}}{2}}
\right)\right].
\end{eqnarray}
Using Eq.(\ref{D1}), the total tunneling rate at horizon $r=r_{+}$ for $\hbar=1$ is derived as
\begin{equation}
\Gamma=\exp\left[ \frac{-2\pi\left[E-\Omega_{H}J-q\left(\frac{q_{e}r+aq_{m}}
{(r^{2}_{+}+a^{2})^{2}}\right)\right]
(r^{2}_{+}+a^{2})}{r_{+}(1-\frac{a^2}{\ell^2})-\frac{2r^3_{+}}{\ell^2}-M-\frac{\alpha
(1-3\omega)r^{-3\omega}_{+}}{2}}\right].
\end{equation}
Thus, by utilizing Boltzmann equation \begin{equation*}\Gamma_{B}=
\exp\left[\frac{E-j\Omega_{H}-q\left(\frac{q_{e}r+aq_{m}}
{(r^{2}_{+}+a^{2})^{2}}\right)}{T_{H}}\right]\end{equation*}
the Hawking temperature $T_{H}$ at the horizon $r_{+}$
can be obtained as
\begin{equation}
T_{H}=\left[\frac{r_{+}(1-\frac{a^2}{\ell^2})-\frac{2r^3_{+}}{\ell^2}
-M-\frac{\alpha(1-3\omega)r^{-3\omega}_{+}}{2}}{2\pi(r^2_{+}+a^2)}\right].
\end{equation}
The above Hawking temperature for the massive particles
case is similar as for massless case for Kerr-Newman-AdS BH with quintessence.
Moreover, the results for spin-down particles are similar
as for spin-up case with the change of sign. The Hawking
temperature is acquired for both cases and we conclude that both
spin-up and spin-down particles are radiate with alike rate.

\section{Quantum Corrections of $T_{H}$}

In this section, we analyze Hawking temperature for massive charged fermions by considering
tunneling procedure incorporating quantum gravitational effects. The modified form of Dirac equation (\ref{C1}) is given
as follows \cite{B3}
\begin{equation}
-\gamma^{0}\left(\iota\partial_{0}+\frac{q}{\hbar}A_{0}\right)
\Psi=\left(\iota\gamma^{i}\partial_{i}+q\frac{\gamma^{i}}
{\hbar}A_{i}+\frac{m}{\hbar}\right)(1+\beta\hbar^{2}
\partial_{j}\partial^{j}-\beta m^{2})\Psi,\label{C8}
\end{equation}
where $i=1, 2, 3$ signifies the spatial coordinates. Moreover,
the correction parameter $\beta$ for the minimal length $M_{f}$ is defined as
$\beta=\frac{\beta_0}{M^{2}_{f}}$ and $m$ is the mass of fermion particles. The Eq.(\ref{C8}) can be rewritten as
\begin{eqnarray}
&&\left[\iota\gamma^{0}\partial_{0}+q\frac{\gamma^{0}}{\hbar}A_{0}+
\iota\gamma^{i}\partial_{i}(1-\beta m^{2})+\iota\gamma^{i}\beta\hbar^{2}
(\partial_{j}\partial^{j})\partial_{i}+q\frac{\gamma^{i}}{\hbar}A_{i}(1+\right.\nonumber\\
&&\left.\beta\hbar^{2}\partial_{j}\partial^{j}-\beta m^{2})+\frac{m}
{\hbar}(1+\beta\hbar^{2}\partial_{j}\partial^{j}-\beta m^{2})\right]\Psi=0.\label{C9}
\end{eqnarray}
Using coordinate transformation, $\varphi=\phi-\Omega$t, where
\begin{equation}
\Omega=\frac{a(\Delta_{\theta}(r^{2}+a^{2})-\Delta_{r})}{(r^{2}+a^{2})^{2}
\Delta_{\theta}-\Delta_{r}a^{2}sin^2\theta},
\end{equation}
the line-element (\ref{B1}) reduces to the following form
\begin{eqnarray}
ds^{2}&=&-\frac{\Delta_{r}\Sigma^{2}}{(r^{2}+a^{2})^{2}\Delta_{\theta}-\Delta_{r}a^{2}sin^2\theta}dt^{2}+
\frac{\Sigma^2}{\Delta_{r}}dr^2+\frac{\Sigma^2}{\Delta_{\theta}}d\theta^2\nonumber\\
&+&\left(\frac{(r^{2}+a^{2})^{2}\Delta_{\theta}-\Delta_{r}a^{2}sin^2\theta}{\Sigma^{2}}
\right)sin^2\theta d\varphi^{2},\nonumber
\end{eqnarray}
which can also be expressed as
\begin{equation}
ds^{2}=-F(r,\theta)dt^{2}+\frac{1}{G(r,\theta)}dr^{2}+
\frac{1}{K(r,\theta)}d\theta^{2}+H(r,\theta)d\varphi^{2}.\label{B3}
\end{equation}
We can choose $\gamma$ matrices in the following form
\begin{eqnarray}
&&\gamma^{t}=\frac{1}{\sqrt{F(r,\theta)}}\left({\begin{array}{cc}\iota & 0\\
 0 & -\iota\\ \end{array}}\right),~~~\gamma^{r}=\sqrt{g(r,\theta)}
 \left({\begin{array}{cc}0 & \sigma^{3}\\
\sigma^{3} & 0\\ \end{array}}\right),\nonumber\\
&&\gamma^{\theta}=\sqrt{K(r,\theta)}\left({\begin{array}{cc}0 & \sigma^{1}\\
\sigma^{1} & 0\\ \end{array}}\right),~\gamma^{\varphi}=
\frac{1}{\sqrt{H(r,\theta)}}\left({\begin{array} {cc}0 &
\sigma^{2}\\\sigma^{2} & 0\\ \end{array}}\right),\label{C10}
\end{eqnarray}
where $\sigma$'s are defined by
Eq.(\ref{A3}). For sake of simplicity, here we discuss the case of spin-up, following the outward radial
trajectory. The wave function of spin-up particles is defined as
\begin{eqnarray}
\Psi_{\uparrow}(t,r,\theta,\varphi)&=&\left[{\begin{array}{cccc}
A(t,r,\theta,\varphi) \\ 0\\B(t,r,\theta,\varphi) \\ 0\\ \end{array}}\right]
\exp\left[\frac{i}{\hbar}I_{\uparrow}(t,r,\theta,\varphi)\right].\label{C11}
\end{eqnarray}
The terms $\hbar$, $\beta$, $\partial A$ and $\partial B$ are considered only
for the first order, i.e., the higher order contributions are ignored.
Using WKB approximation and Eqs.(\ref{C11}) and (\ref{C10}) in Eq.(\ref{C9}),
we can obtain the following set of equations
\begin{eqnarray}
&&-\frac{\iota A}{\sqrt{F(r,\theta)}}(\partial_{t}I)+q\frac{\iota A}
{\sqrt{F(r,\theta)}}A_{t}-\sqrt{G(r,\theta)}(1-\beta m^{2})
(\partial_{r}I)+(1-\beta m^{2})A\nonumber\\
&&+B\beta\sqrt{G(r,\theta)}(\partial_{r}I)\left[G(r,\theta)
(\partial_{r}I)^{2}+
K(r,\theta)(\partial_{\theta}I)^{2}+\frac{1}{H(r,\theta)}
(\partial_{\varphi}I)^{2}\right]\nonumber\\
&&-Am\beta\left[G(r,\theta)(\partial_{r}I)^{2}+
K(r,\theta)(\partial_{\theta}I)^{2}+
\frac{1}{H(r,\theta)}(\partial_{\varphi}I)^{2}\right]=0,\label{C12}
\end{eqnarray}\begin{eqnarray}
&&\frac{\iota B}{\sqrt{F(r,\theta)}}(\partial_{t}I)-\frac{\iota B}
{\sqrt{F(r,\theta)}}A_{t}-A\sqrt{G(r,\theta)}(1-\beta m^{2})
(\partial_{r}I)+A\beta\sqrt{G(r,\theta)}\nonumber\\
&&(\partial_{r}I)\left[G(r,\theta)(\partial_{r}I)^{2}+K(r,\theta)
(\partial_{\theta}I)^{2}+\frac{1}{H(r,\theta)}(\partial_{\varphi}I)^{2}\right]+
Bm(1-\beta m^{2})\nonumber\\
&&-mB\beta\left[G(r,\theta)(\partial_{r}I)^{2}+K(r,\theta)
(\partial_{\theta}I)^{2}+\frac{1}{H(r,\theta)}(\partial_{\varphi}I)^{2}\right]=0,\label{C13}\\
&&B\left[-\sqrt{K(r,\theta)(1-\beta m^{2})}(\partial_{\theta}I)-\frac{\iota(1-\beta m^{2})}
{\sqrt{H(r,\theta)}}(\partial_{\varphi}I)+\beta (\partial_{\theta}I)
\sqrt{K(r,\theta)}\right.\nonumber\\
&&\left.\left\{G(r,\theta)(\partial_{r}I)^{2}+K(r,\theta)
(\partial_{\theta}I)^{2}+\frac{1}{H(r,\theta)}(\partial_{\varphi}I)^{2}\right\}+
\frac{\iota\beta}{\sqrt{H(r,\theta)}}(\partial_{\varphi}I)\right.\nonumber\\
&&\left.\left\{G(r,\theta)(\partial_{r}I)^{2}+K(r,\theta)
(\partial_{\theta}I)^{2}+\frac{1}{H(r,\theta)}(\partial_{\varphi}I)^{2}\right\}\right]=0,\label{C14}\\
&&A\left[-\sqrt{K(r,\theta)}(1-\beta m^{2})(\partial_{\theta}I)-
\frac{\iota(1-\beta m^{2}) }{\sqrt{H(r,\theta)}}(\partial_{\varphi}I)
+\beta(\partial_{\theta}I)\sqrt{K(r,\theta)}\right.\nonumber\\
&&\left.\left\{G(r,\theta)(\partial_{r}I)^{2}+K(r,\theta)(\partial_{\theta}I)^{2}+\frac{1}
{H(r,\theta)}(\partial_{\varphi}I)^{2}\right\}+
\frac{\iota B}{\sqrt{H(r,\theta)}}(\partial_{\varphi}I)\right.\nonumber\\
&&\left.\left\{G(r,\theta)(\partial_{r}I)^{2}+K(r,\theta)(\partial_{\theta}I)^{2}+\frac{1}
{H(r,\theta)}(\partial_{\varphi}I)^{2}\right\}\right]=0.\label{C15}
\end{eqnarray}
The particle's action is given by
\begin{equation}
I=-(E-\Omega j)t+R(r)+\Theta(\theta,\varphi).\label{C16}
\end{equation}
Using Eq.(\ref{C16}) in Eqs.(\ref{C12})-(\ref{C15}) and by focusing on
Eqs.(\ref{C14}) and (\ref{C15}), we observe that they are
similar after dividing by $A$ and $B$, and can be expressed as
\begin{eqnarray}
&&\left\{ \beta G_{r}(r_{+},\theta)R'^{2}+ K(r_{+},\theta) \beta
J_{\theta}^{2}+\beta H(r_{+},\theta)J_{\varphi}^{2} -(1-\beta
m^{2})\right\}\nonumber\\&&\times\left[\sqrt{K(r_{+},\theta)}J_{\theta}
+\iota\frac{1}{\sqrt{H(r_{+},\theta)}}J_{\varphi}\right]=0,\label{C17}
\end{eqnarray}
where $R'=\partial_{r}R$, $J_{\theta}=\partial_{\theta}\Theta$
and $\partial_{\varphi}=\partial_{\varphi}\Theta$. In Eq.(\ref{C17}), $\beta$ indicates the quantum gravity effects so
it cannot be considered as zero, thus the term in large bracket
equals to zero and provide the solution for $\Theta$. Thus, we can write
\begin{equation}
\left[\sqrt{K(r_{+},\theta)}J_{\theta}+\iota
\frac{1}{\sqrt{H(r_{+},\theta)}}J_{\varphi}\right]=0.\label{C18}
\end{equation}
After removing $A$ and $B$ from Eqs.(\ref{C12}) and (\ref{C13}), we
get the following expression
\begin{equation}
P_{6}(\partial_{r}R)^{6}+P_{4}(\partial_{r}R)^{4}
+P_{2}(\partial_{r}R)^{2}+P_{0}=0,\label{A4}
\end{equation}
where
\begin{eqnarray}
&&P_{6}=\beta^{2}G^{3}F,\nonumber\\
&&P_{4}=\beta G^{2}F(m^{2}\beta+2\beta C-2),\nonumber\\
&&P_{2}=GF\left[(1-\beta m^{2})^{2}+\beta(2m^{2}-2m^{4}\beta-2C+\beta C^{2})\right],\nonumber\\
&&P_{0}=m^{2}(1-\beta m^{2}-\beta C)^{2}F-(E-\Omega j+qA_{t})^{2},\nonumber\\
&&C=K(r_{+},\theta)J_{\theta}^{2}+\frac{1}{H(r_{+},\theta)}J_{\varphi}^{2}.\nonumber
\end{eqnarray}
From Eq.(\ref{C18}), we note that $C=0$. Considering $\beta$ only
for the first order, the solution of Eq.(\ref{A4}) at horizon provides \cite{C3}
\begin{eqnarray}
&&R(r)=\pm\int\frac{1}{\sqrt{GF}}\sqrt{m^{2}(1-2\beta m^{2})F
+(E-\Omega j-qA_{t})^{2}}\times\nonumber\\&&\left[1+\beta
\left(m^{2}+\frac{(E-\Omega j-qA_{t})^{2}}{F}\right)\right]dr.\nonumber
\end{eqnarray}
The above equation implies
\begin{equation}
R(r)=\pm\iota\pi\frac{(r^{2}_{+}+a^{2})(E-\Omega_{H} j-qA_{t})}
{\Delta_{r}(r_{+})}(1+\beta\Xi),
\end{equation}
where
\begin{equation*}
\Xi=6m^{2}+\frac{6}{r^{2}+a^{2}\cos^{2}\theta}
\left(J^{2}_{\theta}+J^{2}_{\varphi}\csc^{2}\theta\right).
\end{equation*}
The positive/negative signs indicate outgoing/incoming particles. Hence, the tunneling rate \cite{C4} of charged
fermions across the horizon is calculated as
\begin{eqnarray}
\Gamma&=&\frac{\Gamma_{(emission)}}{\Gamma_{(absorption)}}=\frac{\exp[-\frac{2}{\hbar}
(ImR_{+}+Im\Theta)]}{\exp[-\frac{2}{\hbar}(ImR_{-}+Im\Theta)]}=
\exp[-\frac{4}{\hbar}ImR_{+}],\nonumber\\&=&\exp\left[\frac{-2\pi
\left\{E-\Omega_{H}j-q\left(\frac{q_{e}r+aq_{m}}{(r^{2}_{+}+a^{2})^{2}}\right)\right\}
(r^{2}_{+}+a^{2})}{r_{+}(1-\frac{a^2}{\ell^2})-\frac{2r^3_{+}}{\ell^2}-M-
\frac{\alpha(1-3\omega)r^{-3\omega}_{+}}{2}}\right](1+\beta\Xi).\nonumber
\end{eqnarray}
For $\hbar=1$, the corrected Hawking temperature $T_{e-H}$ is
\begin{eqnarray}
T_{e-H}&=&\left[\frac{r_{+}(1-\frac{a^2}{\ell^2})-
\frac{2r^3_{+}}{\ell^2}-M-\frac{\alpha(1-3\omega)
r^{-3\omega}_{+}}{2}}{2\pi(r^2_{+}+a^2)}\right](1+\beta\Xi)^{-1},\nonumber\\
&=&T_{\circ}\left[1-\beta\Xi+(\beta\Xi)^2+...\right],\label{WM1}\\
&\approx&T_{\circ}(1-\beta\Xi).\label{C19}
\end{eqnarray}
where the semi-classical Hawking temperature is
\begin{equation*}
T_{\circ}=\left[\frac{r_{+}(1-\frac{a^2}{\ell^2})-\frac{2r^3_{+}}{\ell^2}-M-
\frac{\alpha(1-3\omega)r^{-3\omega}_{+}}{2}}{2\pi(r^2_{+}+a^2)}\right].
\end{equation*}
The corrected temperature is based on quantum numbers, i.e., mass,
energy and angular momentum.

\section{Logarithmic Entropy Corrections}

This section is devoted to calculate the entropy corrections for Kerr-Newman-AdS BH
with quintessence. Using null geodesic technique, Banerjee and Majhi \cite{S1} investigated the
corrected Hawking temperature and corrected entropy by taking into account the back-reaction effects.
Majhi \cite{S3} also analyzed the corrected temperature and entropy by using the
first and second laws of thermodynamics. In our investigation, we calculate the
entropy corrections for Kerr-Newman-AdS BH involving quintessence by considering
the generic formula for leading order corrections to Bekenstein-Hawking formula \cite{D5}.
It is worth mentioning here that, one could calculate the logarithmic
corrections to the BH entropy at inner/outer horizon ($r_{\pm}$) without
knowing the values of any specific heat of the BH but only knowing the
values of Hawking temperature $(T_{{e-H},{\pm}})$ and entropy ($S_{0,\pm}$),
for the given BH.
The BH entropy corrections ($S_{\pm}$) can be defined as
\begin{equation}
S_{\pm}=S_{0, \pm}-\frac{1}{2}\ln\mid T^2_{e-H,\pm}S_{0,\pm}\mid+...~.\label{z1}
\end{equation}
The classical entropy for given BH at $r=r_+$ can be calculated as follows
\begin{equation}
S_{0, +}=\frac{\mathcal{A}_+}{4}=\frac{\pi\left(r_{+}^2+a^2\right)}{\left(1-\frac{a^2}{\ell^2}\right)},
\end{equation}
where
\begin{equation*}
\mathcal{A}_{+}=\int_{0}^{2\pi}\int_{0}^{\pi}\sqrt{g_{\theta\theta}g_{\phi\phi}d\theta d\phi}.
\end{equation*}
After substituting the corrected Hawking temperature $T_{e-H}$ (given by Eq.(\ref{C19})) in the
above Eq.(\ref{z1}), we can obtain the logarithmic corrections of entropy in the following form
\begin{eqnarray}
S=S_{0, +}-\frac{1}{2}\ln\left|\frac{\left(r_{+}(1-\frac{a^2}{\ell^2})-
\frac{2r^3_{+}}{\ell^2}-M-\frac{\alpha(1-3\omega)
r^{-3\omega}_{+}}{2}\right)^2}{4\pi(r^2_{+}+a^2)\left(1-\frac{a^2}{\ell^2}\right)}(1-\beta\Xi)^2\right|+...\nonumber\\
\end{eqnarray}
This expression represents the corrected entropy for Kerr-Newman-AdS BH involving quintessence parameter.

\section{Effects of $\beta$ on $T_{e-H}$}

In this section, we analyze graphically the effects of $\beta$ on $T_{e-H}$ with respect to
various parameters.

\subsection{Temperature $T_{e-H}$ with Horizon $r_{+}$ }

This subsection is devoted to discuss the behavior of corrected Hawking
temperature in different domains of event Horizon $r_{+}$ with fixed BH
mass $M=1$, state parameter $\omega=-\frac{2}{3}$ and $\Xi=0.01$.
\begin{figure}
\epsfig{file=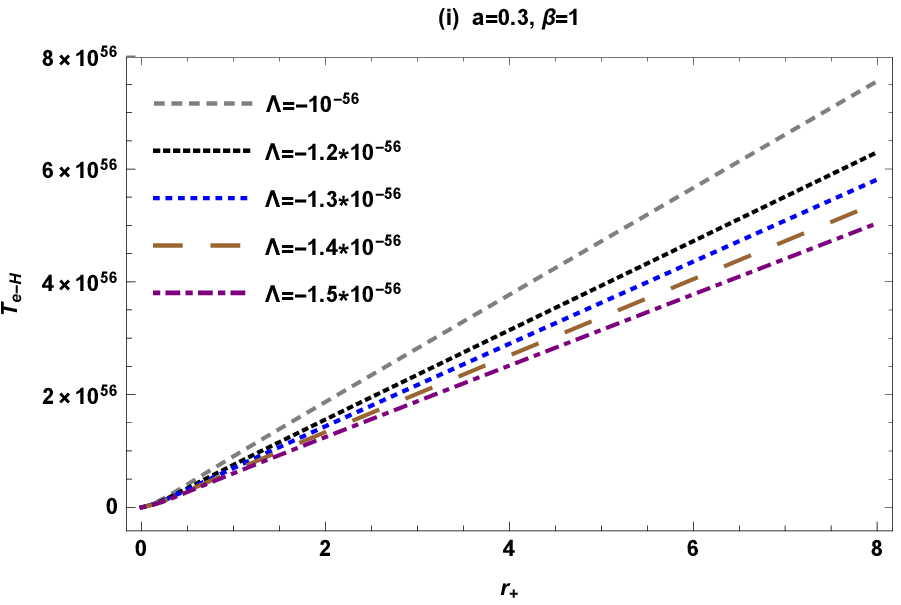,width=0.5\linewidth}\epsfig{file=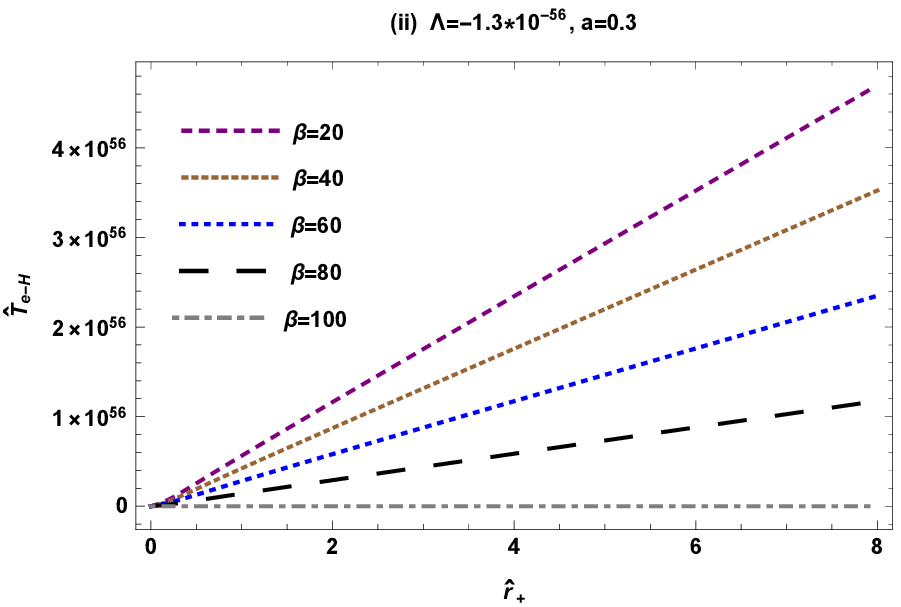,width=0.5\linewidth}
\caption{For $\alpha=0.01$, Hawking temperature $T_{e-H}$ with
respect to horizon $r_{+}$.}
\epsfig{file=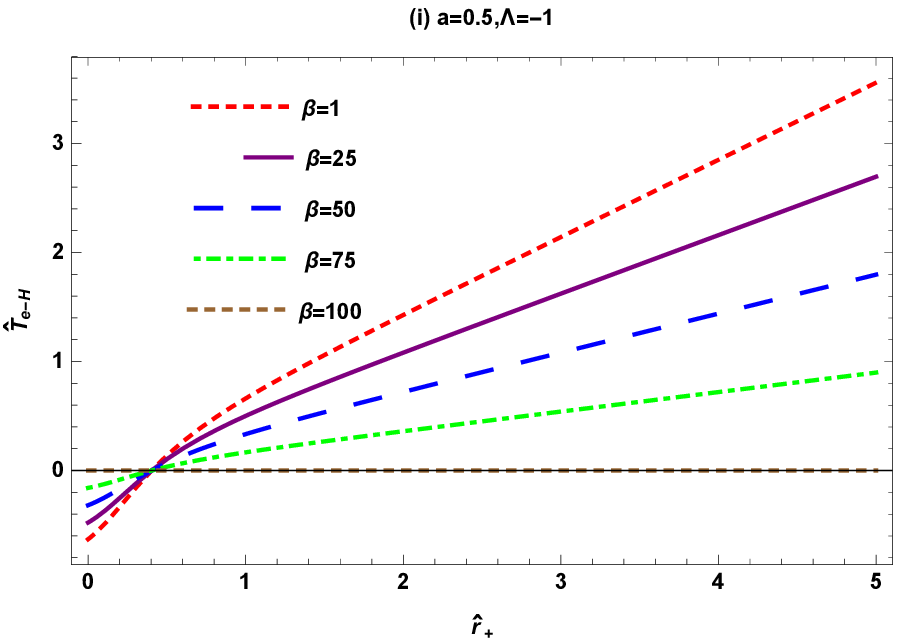,width=0.5\linewidth}\epsfig{file=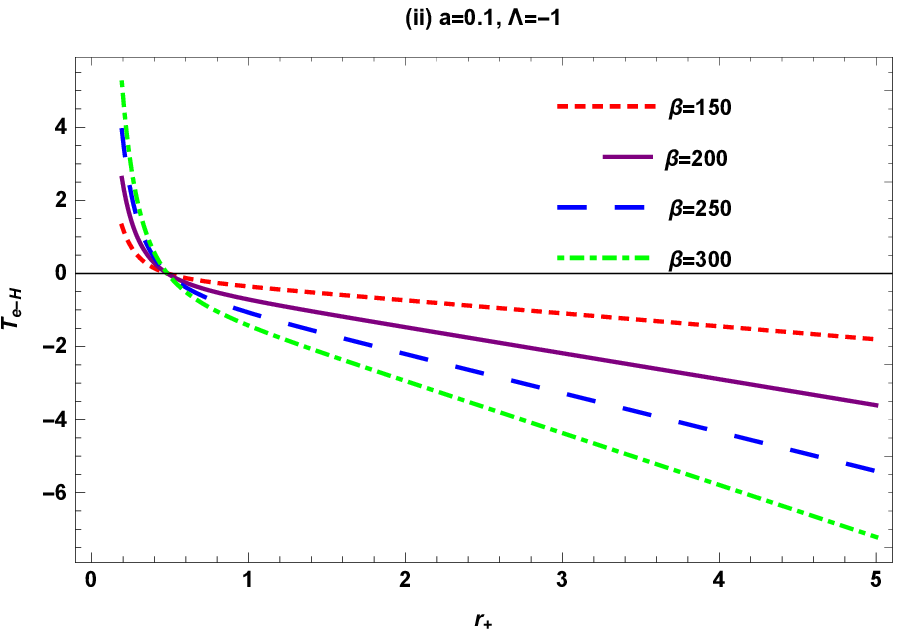,width=0.5\linewidth}
\caption{For $\alpha=1$, Relation between the temperature and $r_{+}$.}
\epsfig{file=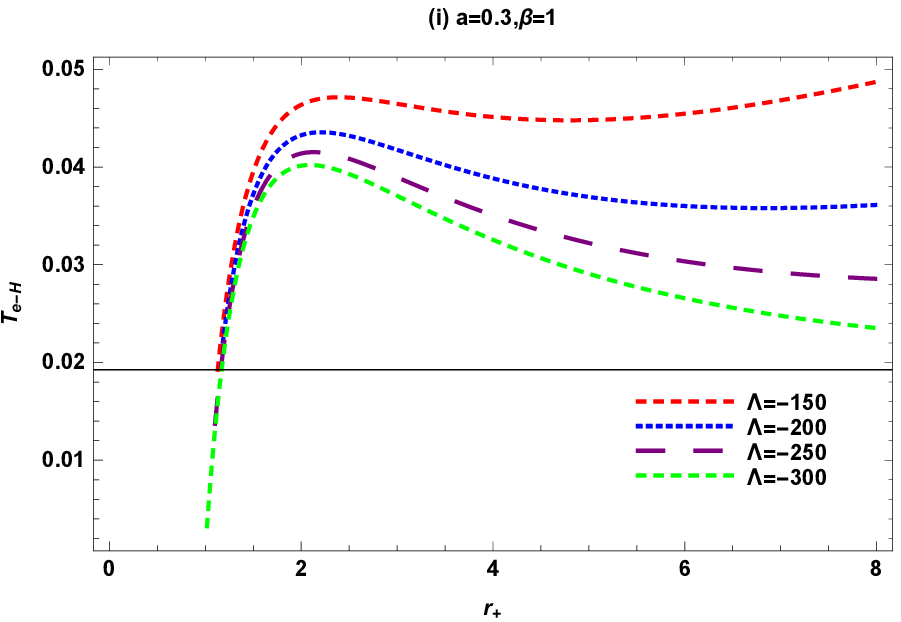,width=0.5\linewidth}\epsfig{file=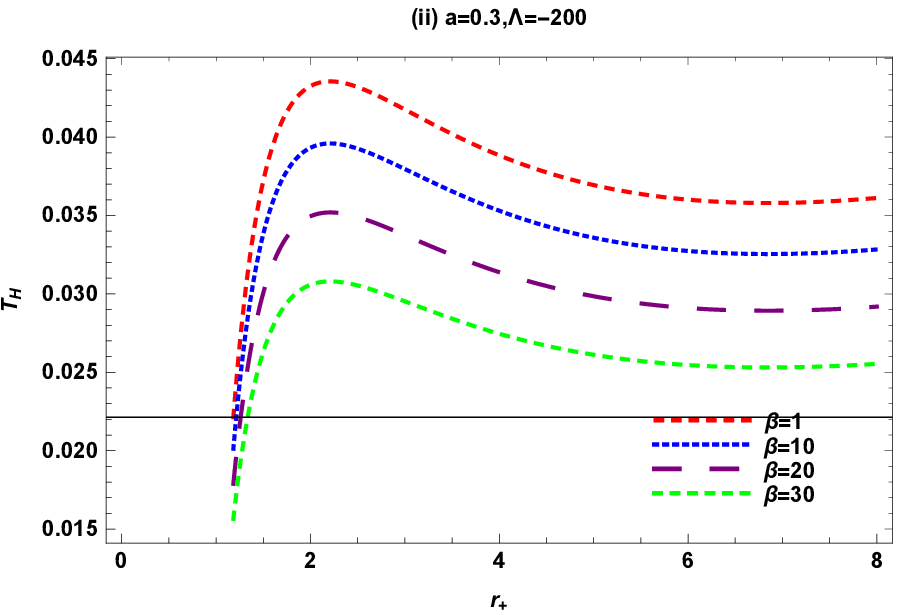,width=0.5\linewidth}
\caption{For $\alpha=1$, Relation between the temperature and
$r_{+}$.}
\end{figure}
\begin{itemize}
\item \textbf{Figure 1} indicates the behavior of Hawking temperature with $r_+>0$, for
different values of cosmological constant $\Lambda<0$ with fixed values of rotation
parameter $a=0.3$ and correction parameter
$\beta=1$. The cosmic
repulsion indicates that the recent value of the cosmological
constant is $\Lambda\approx-1.3\times10^{-56} cm^{-2}$ \cite{B1}.
In our analysis, we consider this fixed value of $\Lambda=-1.3\times10^{-56}
cm^{-2}$, as well as we consider $\Lambda$ greater and lesser than
this fixed value. \textbf{(i)}: For different values of $\Lambda$ near to the fixed value,
we can observe that the temperature has its maximum value and its
behavior is linear. As horizon increases the temperature also
increases. The change in $\Lambda$ defines the diverging temperature
$T_{e-H}$ as $r_{+}$ increases. \textbf{(ii)}: Indicates the
behavior of temperature for fixed values of
$a=0.3$ and $\Lambda=-1.3\times10^{-56}$, while the
values of $\beta$ varies. We can observe that
for $\beta<100$, the temperature is going to increase as $\beta$ going lesser from $100$. While, for $\beta=100$, the temperature will be zero. The temperature has linear behavior, it increases as horizon increases.
\item \textbf{Figure 2} indicates the behavior of $T_{e-H}$ for fixed values of
$a$ and $\Lambda$, while for varying $\beta$.  \textbf{(i)}: We can observe that for fixed
value of $a=0.5$, $\Lambda=-1$ and for varying $\beta<100$, the
behavior of temperature is positively increasing $T_{e-H}>0$. While, for $\beta=100$, the graph shows \textit{zero} temperature, i.e.,
$T_{e-H}=0$.
\textbf{(ii)}: For fixed $a=0.1$, $\Lambda=-1$ and for varying $\beta>100$,
we observe the negatively divergent behavior of temperature, i.e., the temperature is decreasing as
horizon increases in the negative range. This negative and divergent behavior
of temperature reflects the non-physical unstable state of BH.

It is also worth mentioning here that, for the values of correction parameter $1 \leq \beta <100$,
we observe positive values of temperature and for $\beta=100$, the temperature vanishes, while for $\beta>100$,
we observe non-physical behavior with negative temperature.
\item\textbf{Figure 3} shows the behavior of temperature
for fixed values of $a=0.3$ and varying $\Lambda$ and $\beta$.
\textbf{(i)}: We can observe that for $a=0.3$ and
$\beta=1$, as $\Lambda$ decreases, the temperature will be small
and after attaining its maximum value, the temperature decreases.
For $\Lambda\ll0$, the temperature decreases as horizon increases.
In all curves, the temperature has its maximum value at small
horizon. The horizon of BH can never be zero, the non-zero value of
horizon leads to a BH remnant with maximum temperature.\textbf{(ii)}: We observe that for $a=0.3$,
$\Lambda=-200$ and for varying $\beta<100$, the temperature attains positive values, Initially, the temperature will be maximum at non-zero horizon and later decreases exponentially as horizon increases, which indicates
a physical stable state of BH.
\end{itemize}
\begin{figure}
\epsfig{file=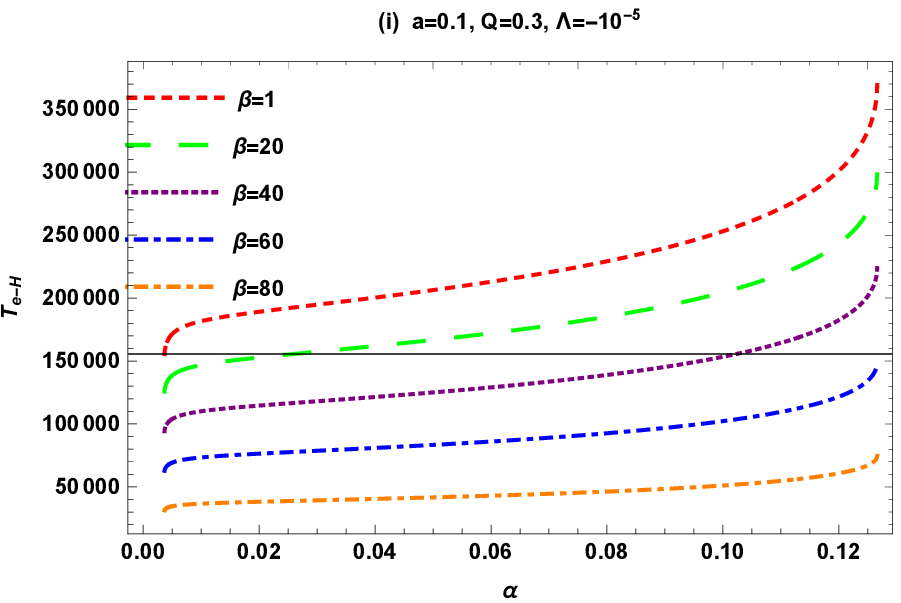,width=0.5\linewidth}
\epsfig{file=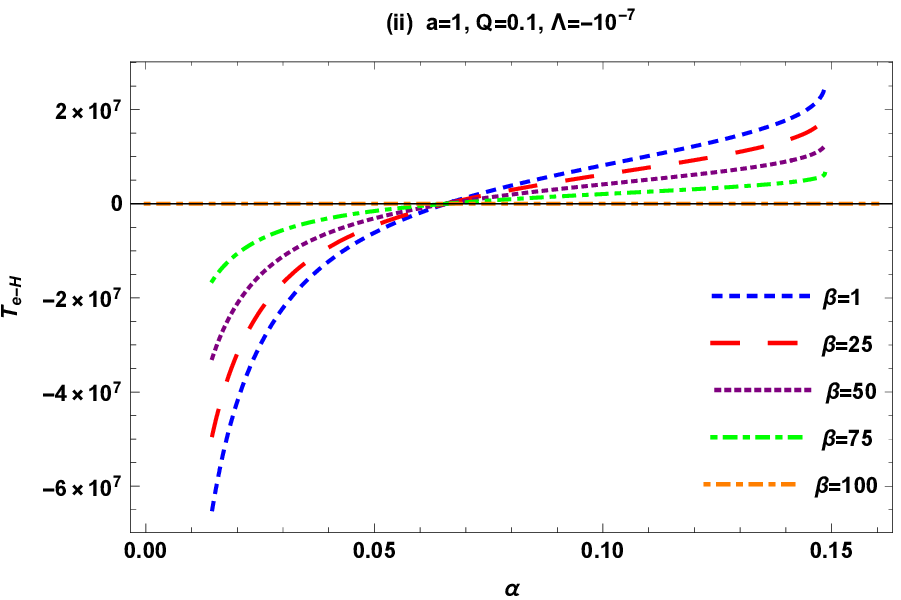,width=0.5\linewidth}\caption{Relation between
temperature and $\alpha$ in term of $\beta$.}
\epsfig{file=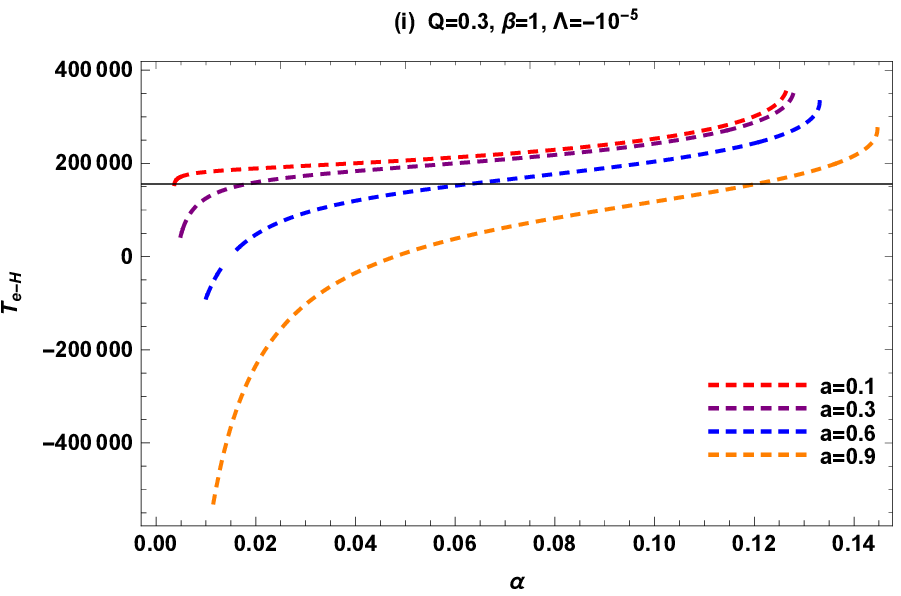,width=0.5\linewidth}\epsfig{file=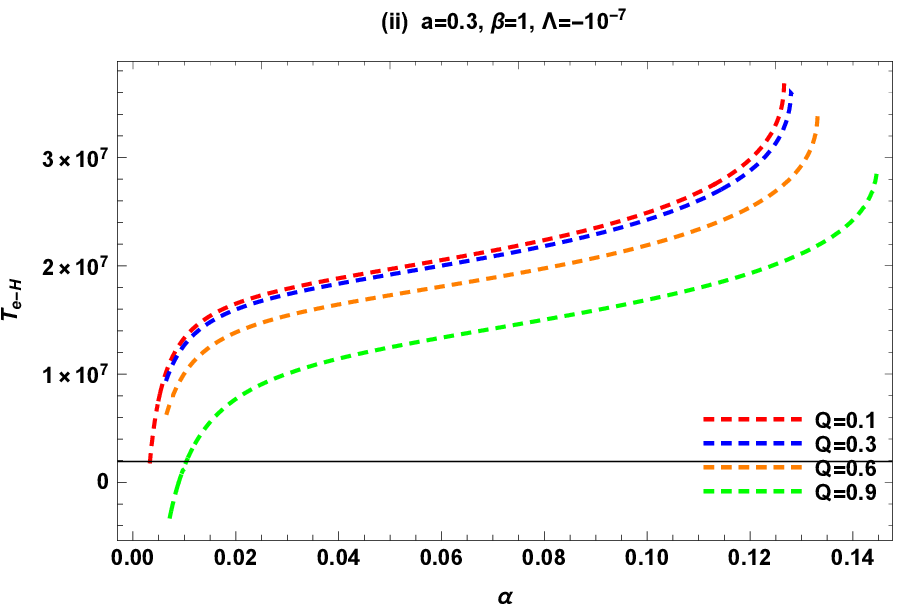,width=0.5\linewidth}\caption{Relation
between temperature and $\alpha$ in term of $a$ and $Q$.}
\end{figure}

\subsection{Temperature $T_{e-H}$ with Quintessence $\alpha$}

This subsection gives the analysis of corrected Hawking temperature $T_{e-H}$ with
quintessence parameter $\alpha$ for different values of rotation
parameters $a$, BH charge $Q$ and cosmological constant $\Lambda$
for fixed $M=1$, $\omega=-\frac{2}{3}$ and $\Xi=0.01$.
\begin{itemize}
\item \textbf{Figure 4} shows the behavior of temperature for fixed
$a$, $Q$, $\Lambda$ and for varying $\beta$. \textbf{(i)}: We can
observe that for fixed values of $a=0.1$, $Q=0.3$,
$\Lambda=-10^{-5}$ and for different values of $\beta$, the
temperature gradually increases as $\beta$ decreases. It is to be noted that
for these parameters, the temperature increases
with increase in $\alpha$ as $\beta\rightarrow1$. \textbf{(ii)}: This graph indicates the behavior of
temperature for fixed values of $a=1$, $Q=0.1$, $\Lambda=-10^{-7}$
and for varying $\beta$. It is to be noted that for $\beta<100$, the
behavior of temperature is from negative to positive, while it attains maximum values and temperature increases with increase in $\alpha$.
While, for $\beta=100$, the temperature vanishes, i.e., $T_{e-H}=0$.
It is important to note that the both negative and
positive behaviors of temperature shows that the initial unstable state of BH, which turns out to be stable with time. This negative temperature is the effect of rotation parameter $a=1$ (maximum value).
\item \textbf{Figure 5} shows the behavior of temperature for fixed $\Lambda$ and $\beta$, while for
varying $a$ (in Fig.\textbf{(i)}) and varying $Q$ (in Fig.\textbf{(ii)}). \textbf{(i)}: We can
observe that for fixed values of $Q=0.3$, $\beta=1$,
$\Lambda=-10^{-5}$ and for varying $a$, the temperature gradually
increases for increasing
$\alpha$. It is to be noted that as
$a\rightarrow1$, the temperature will goes on from negative
to positive. \textbf{(ii)}: We can
observe that for fixed values of $a=0.3$, $\beta=1$,
$\Lambda=-10^{-7}$ and for varying $Q$, the temperature will be high enough and it will gradually
increase with increase in $\alpha$.
\end{itemize}

\subsection{3D Plots: $T_{e-H}$ and $r_{+}$ with $\Lambda$, $\alpha$ and $\beta$ }

This section is based on the analysis of Hawking temperature with
quintessence parameter $\alpha$, horizon $r_{+}$ and cosmological
constant $\Lambda$ for fixed $M=1$, $\omega=-\frac{2}{3}$ and
$\Xi=0.01$. This section consists of $3D$ graphs.
\begin{figure}\begin{center}
\epsfig{file=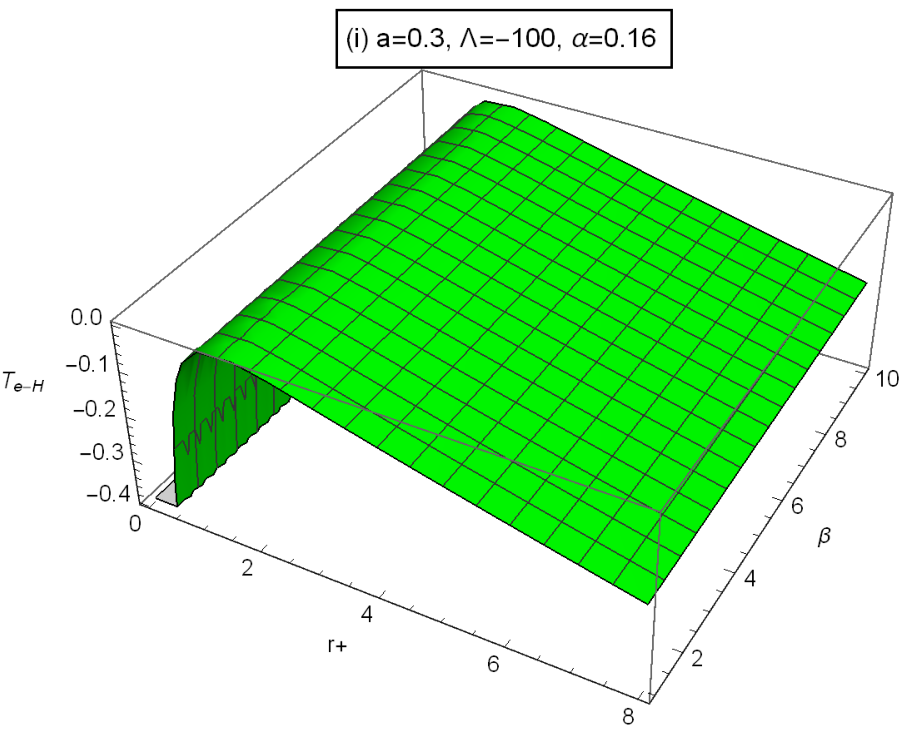,width=0.47\linewidth}\epsfig{file=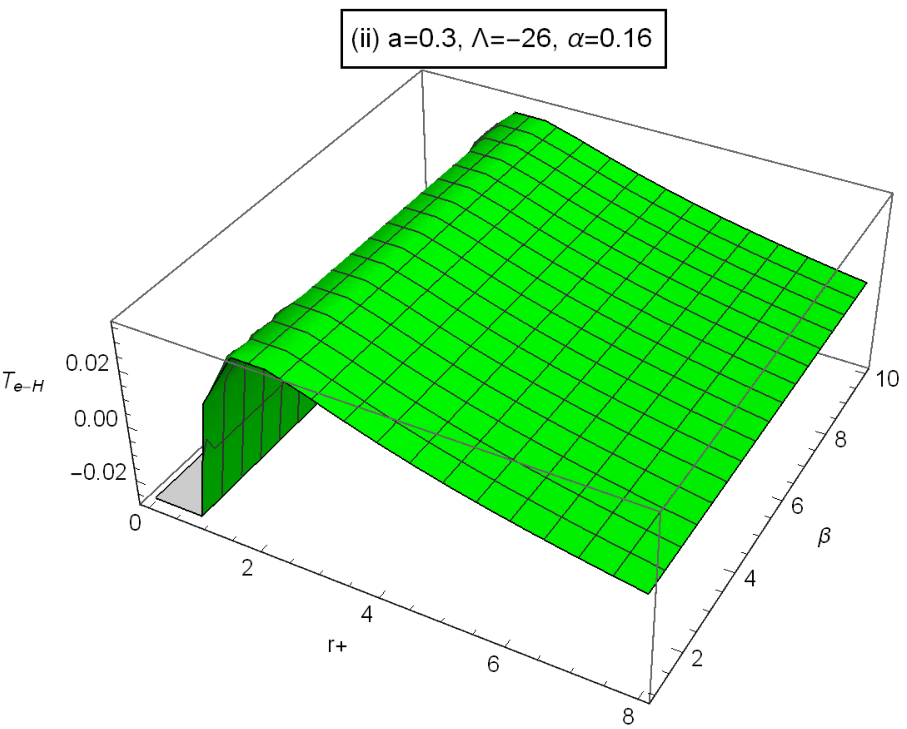,width=0.47\linewidth}
\caption{Relation between temperature with $r_{+}$ and $\beta$.}
\epsfig{file=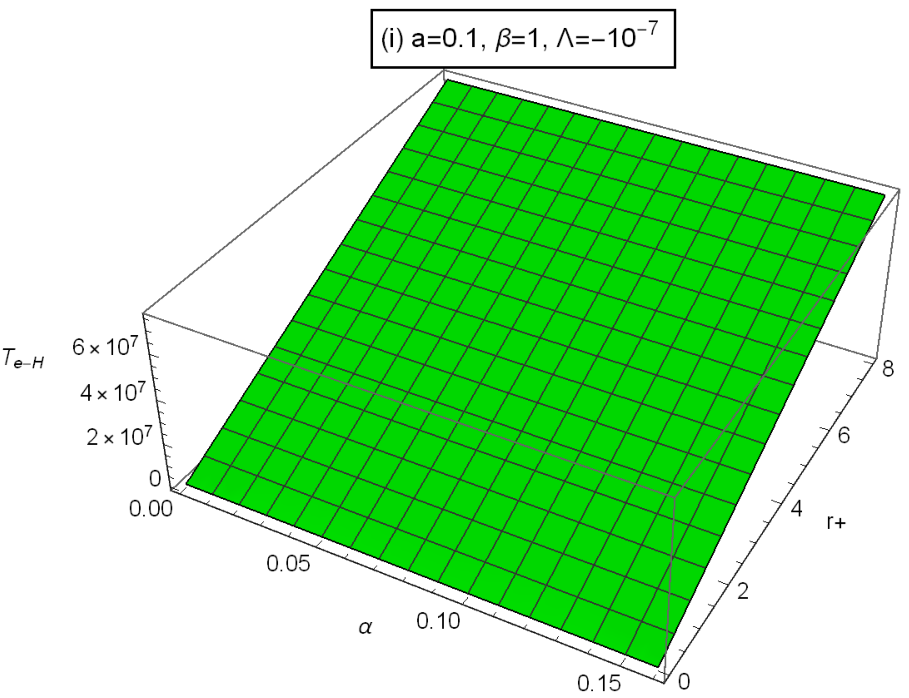,width=0.47\linewidth}\epsfig{file=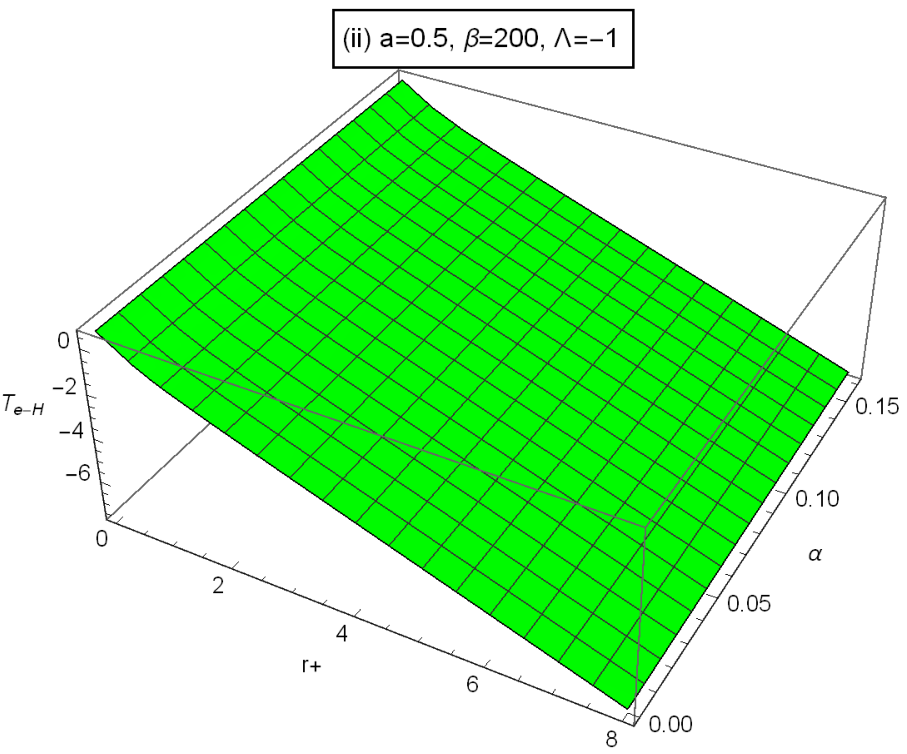,width=0.47\linewidth}
\caption{Relation between temperature with $r_{+}$ and $\alpha$.}
\epsfig{file=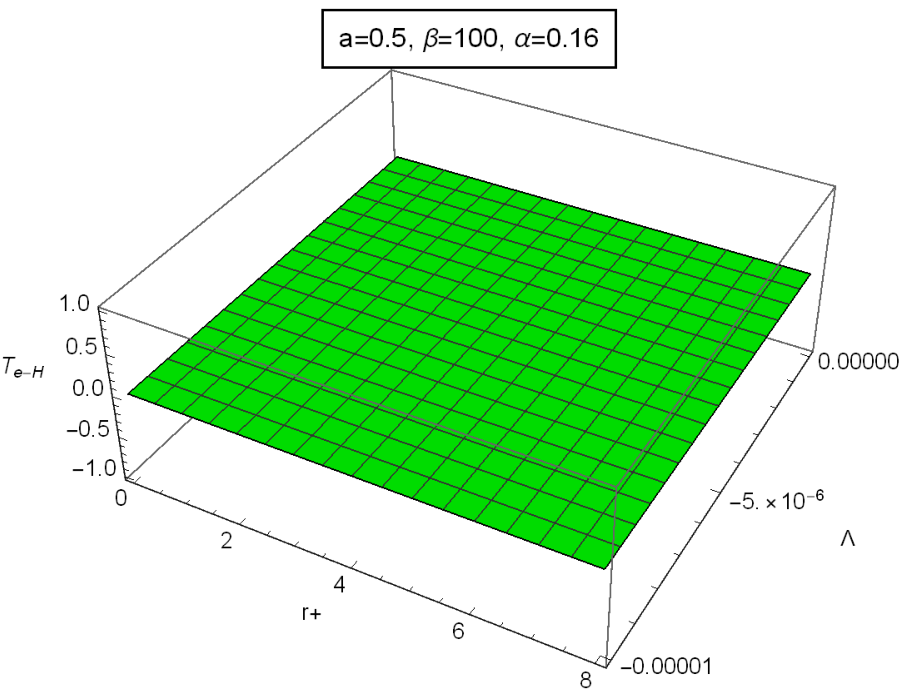,width=0.47\linewidth}
\caption{Relation between temperature with $r_{+}$ and $\Lambda$.}
\end{center}
\end{figure}
\begin{itemize}
\item \textbf{Figure 6} shows the behavior of temperature for fixed $a$,
$\Lambda$ and $\alpha$, while for $\beta<100$. \textbf{(i)}: Here, for fixed $a=0.3$,
$\Lambda=-100$ and $\alpha=0.16$, the behavior of $T_{e-H}$ indicates that the
temperature attains negative values and its behavior is constant w.r.t. $\beta$. After
attaining its maximum value at non-zero horizon, the Hawking
temperature decreases as horizon increases. This indicate that
there exist a BH remnant during the evaporation process. The BH
temperature attains its maximum value as horizon shrinks. For $\Lambda=-100$, the temperature is negative for
$\beta<100$ and invert for other $\beta$ (already mentioned in $2D$ graphs0. \textbf{(ii)}:
Graph indicates the behavior of temperature for fixed $a=0.3$,
$\Lambda=-26$ and $\alpha=0.16$. We can observe that the behavior of
temperature is same as in \textbf{(i)}. For $\Lambda=-100$, the
temperature is negative but as we increase the value of $\Lambda$
till $\Lambda=-26$, the graph will exhibit the behavior of
temperature from negative to positive and for $\Lambda=-25$, the
graph will exhibit the positive temperature.
\item \textbf{Figure 7} shows the behavior of $T_{e-H}$ with
$r_{+}$ and $\alpha$ for fixed $a$, $\beta$ and $\Lambda$.
\textbf{(i)}: Graph indicates the behavior of $T_{e-H}$ for fixed
$a=0.1$, $\beta=1$ and $\Lambda=-10^{-7}$. We can observe that for
larger value of $\Lambda$, the behavior of temperature is linear
w.r.t. $r_{+}$. The temperature attains high values and increases
with increase in $\alpha$. \textbf{(ii)}: This figure indicates the
behavior of temperature for fixed $a=0.5$, $\beta=200$ and
$\Lambda=-1$. We can observe that as horizon increases the
temperature decreases. While for $\alpha$, the temperature increases
as $\alpha$ increases. In \textbf{Fig. 7}, the behavior of
temperature is same as proved in $2D$ graphs.
\item \textbf{Figure 8} shows the behavior of temperature with
horizon $r_{+}$ and cosmological constant $\Lambda$ for fixed
$a=0.5$, $\beta=100$ and $\alpha=0.16$. We can observe that for
$\beta=100$, the effect of temperature will be zero, as concluded in
$2D$ graphs.
\end{itemize}
It is worth mentioning here that from 2D and 3D graphs, we conclude that the
temperature increases with decreasing horizon
and increasing $\alpha$. For small values of $\Lambda\neq-100$ and
$\beta>100$, the temperature shows negative values. For larger values of
$\Lambda$, the temperature will be maximum and its behavior is
linear w.r.t. horizon. Moreover, for large $\Lambda$, the
temperature increases with increasing $\alpha$.

\section{Conclusion}

In this paper, we have computed radiation spectrum by analyzing Hawking temperature for
Kerr-Newman-AdS BH surrounded by quintessence. For this purpose, we have
utilized the WKB approximation and the Hamilton-Jacobi ansatz for
massive charged spin-$\frac{1}{2}$ particles (fermions). The investigation
yields the corrected Hawking temperature $T_{e-H}$, reliable with BH universality.
In our analysis, we have altered the Dirac equation in curved spacetime by
incorporating quantum gravity effects through GUP. We have evaluated the tunneling
rate at horizon as well as the corresponding Hawking temperature, quantum corrected
Hawking temperature as well as quantum corrected entropy. We have analyzed in detail
quantum corrected Hawking temperature graphically.

We have summarized the detailed graphical analysis of this paper in the following points:
\begin{itemize}
\item The derived Hawking temperature and its modified form $T_{e-H}$ depends
on BH's mass, charge and rotation parameters, as well as on the mass and angular momentum of the emitted
fermion particles, quintessence parameter $\alpha$ and state parameter $\omega$.

\item When the quantum gravity effects are neglected ($\beta=0$), we have recovered
the Hawking temperature of Kerr Newman AdS BH with
quintessence. For $\alpha=0$ and $\omega=0$, we have obtained the Hawking temperature
of Kerr-Newman AdS BH. Moreover, when $\Lambda=0$, the temperature of
Kerr Newman BH has obtained. In addition, for charge-free ($Q=0$) and non-rotating ($a=0$)
case, the temperature and its correction reduce to the case of Schwarzschild BH.

\item In our analysis, we have considered $\Xi=0.01$, then the condition of GUP must be satisfied
for $0<\beta<100$. We have substituted the above mentioned values in Eq.(\ref{WM1})
for positive values of $T_\circ$, the correction terms became smaller than the
previous terms given in series. When we consider the first order quantum corrections,
the correction term is smaller than $T_\circ$. For $\beta=100$, the first order
correction term is same as semi-classical term $T_\circ$, showed invalidity of
GUP.  When $\beta>100$, the correction term became greater than the preceding
term and the condition of GUP is not satisfied.

\item Graphical analysis showed that the behavior of $T_{e-H}$ is positive and
negative when $\beta<100$ and $\beta>100$, respectively. While, for $\beta=100$, the temperature vanishes.

\item For fixed $\Lambda\approx-1.3\times10^{-56}$ \cite{B1},
as well as greater and lesser than this fixed value of $\Lambda$,
we have observed that the temperature increases as horizon increases,
which is non-physical. The negative as well as divergent behavior of Hawking temperature indicates the behavior reverse to Hawking's phenomenon,
representing unstable state of BH. While, for smaller $\Lambda\ll0$, the temperature decreases with increasing horizon, which is physical.

\item For smaller $\Lambda$, we have obtained physical (stable, \textit{+ive} $T_{e-H}$) and non-physical (\textit{-ive} $T_{e-H}$) behavior of temperature for $\beta<100$ and $\beta>100$, respectively.

\item We have observed the behavior of $T_{e-H}$ w.r.t.
$\alpha$ only for the particular ranges, i.e., $0<\alpha<1/6$, $0<Q\leq1$, $\Lambda\leq-1$
and $0<a\leq1$.

\item For $T_{e-H}$ w.r.t.
$\alpha$, at maximum value of the rotation parameter (i.e., $a=1$), we have observed non-physical and
unstable state of BH, which is due to the instability in temperature.

\item The results obtained from 3D graphs are similar to the results obtained from 2D graphs.
\end{itemize}

\end{document}